\documentclass[twocolumn,aps,prd,amsmath,superscriptaddress]{revtex4-1}
\usepackage{cancel}
\usepackage{graphicx}
\usepackage{color}

\begin{document}

\title{Compact stars in a SU(3) Quark-Meson Model}

\author{Andreas Zacchi}
\affiliation{Institut f\"ur Theoretische Physik, Goethe Universit\"at,
Max-von-Laue-Stra\ss{}e 1, D-60438 Frankfurt, Germany} 

\author{Rainer Stiele}
\affiliation{Institut f\"ur Theoretische Physik, Goethe Universit\"at,
Max-von-Laue-Stra\ss{}e 1, D-60438 Frankfurt, Germany} 
\affiliation{Institut f\"ur Theoretische Physik, Universit\"at Heidelberg,
Philosophenweg 16, D-69120 Heidelberg, Germany}

\author{J\"urgen Schaffner-Bielich}
\affiliation{Institut f\"ur Theoretische Physik, Goethe Universit\"at,
Max-von-Laue-Stra\ss{}e 1, D-60438 Frankfurt, Germany} 

\date{\today}

\begin{abstract}
  The recent observations of the massive pulsars PSR J1614-2230 and of PSR
  J0348+0432 with about two solar masses imply strong constraints on the
  properties of dense matter in the core of compact stars.  Effective models
  of QCD aiming to describe neutron star matter can thereby be considerably
  constrained. In this context, a chiral quark-meson model based on a SU(3)
  linear $\sigma$-model with a vacuum pressure and vector meson exchange is
  discussed in this work. The impact of its various terms and parameters on
  the equation of state and the maximum mass of compact stars are delineated
  to check whether pure quark stars with two solar masses are feasible within
  this approach. Large vector meson coupling constant and a small vacuum
  pressure allow for maximum masses of two or more solar masses. However,
  pure quark stars made of absolutely stable strange quark matter, so called
  strange stars, turn out to be restricted to a quite small parameter range.
\end{abstract}

\maketitle

\section{Introduction}

The core of a massive star which has exhausted its thermonuclear fuel
collapses in a supernova which is one of the most extreme events known to
occur in the universe. The relic in the aftermath of such a cataclysmic event
constitutes also one of the most extreme objects known to exist in the
universe.  Being more massive than our sun, but only around 20 to 30
kilometers in diameter, these so called neutron stars harbor the densest
material known at present.  The density in the core of such an object can even exceed 
nuclear density ($\rho_0 \approx 2.5 \cdot 10^{14}$ g/cm$^3$).

For many decades it has been speculated that the material in the core of
neutron stars might consist of quark matter
\cite{Ivanenko:1965dg,Itoh:1970uw}.  Neutron stars, or better compact stars,
with a core consisting of quark matter are dubbed hybrid stars. Compact stars
which are entirely made of quark matter, besides maybe a small layer of a crust of
nuclei, are so called strange stars
\cite{Haensel:1986qb,Alcock:1986hz}. Strange stars can only be realized in
nature if strange quark matter is absolutely stable, i.e.\ the true ground
state of matter \cite{Bodmer:1971we,Witten:1984rs}. Hybrid stars as well as
strange stars have been usually modeled by using the MIT bag model, see
\cite{Weber:2004kj} for a review, also with corrections from hard-dense-loop
calculations \cite{Schertler:1997za}.  The Nambu-Jona-Lasinio (NJL) model has
been adopted for describing quark matter in compact stars by using scalar
meson fields only in \cite{Schertler:1999xn,Steiner:2000bi}. Vector meson
fields have been added within the NJL-model in \cite{Hanauske:2001nc} showing
that their contribution can substantially increase the maximum mass of a
compact star with quark matter. Effects from color-superconductivity have been
also considered for the properties of compact stars within NJL-type models,
see \cite{Buballa:2003qv} for a review.  We point out that it was
well known that hybrid stars can be as massive as two solar masses in various
approaches \cite{Klahn:2006iw,Alford:2006vz} and could even masquerade as
neutron stars \cite{Alford:2004pf}.

The recent measurements of the masses of the pulsars PSR J1614-2230
\cite{Demorest:2010bx} and of PSR J0348+0432 \cite{Antoniadis:2013pzd} with
$M=2.01\pm 0.04 M_{\odot}$ impose considerable constraints on the equation of
state (EoS) for compact stars. Pure quark stars based on the simple MIT bag
model could be ruled out on the basis of this measurement unless additional
terms from an effective one-gluon exchange or from color-superconductivity are
incorporated \cite{Ozel:2010bz,Weissenborn:2011qu}. However, the MIT bag model
fails in describing QCD lattice data at nonvanishing temperature questioning
its applicability for describing dense quark matter \cite{Fraga:2013qra}.
Effective models of QCD based on chiral symmetry are able to describe the
lattice data at nonvanishing temperatures, as the Polyakov-loop extended
versions of the NJL-model \cite{Ratti:2005jh} or the Polyakov-loop quark-meson
(PQM) model \cite{Schaefer:2007pw,Haas:2013qwp,Herbst:2013ufa}. NJL-type models for hybrid
stars have been investigated for being compatible with a maximum mass of
compact stars of at least two solar masses by several groups
\cite{Blaschke:2010vd,Bonanno:2011ch,Masuda:2012ed,Masuda:2012kf,
  Orsaria:2012je,Orsaria:2013hna,Benic:2014iaa,Yasutake:2014oxa,Benic:2014jia}. 
Recently, the connection between NJL-type
models, the Schwinger-Dyson approach and an extended version of the MIT bag
model with vector-like interaction terms has been pointed out in
\cite{Klahn:2015mfa,Chen:2011my,Chen:2015mda} for modeling compact star matter incorporating features
of QCD.  To our knowledge, the quark-meson model has not been used to
investigate the properties of hybrid stars or quark stars so far, in
particular not in view of the now well established two solar mass limit for
compact stars.

In this work, a modified linear-$\sigma$-model is used to describe compact
star matter consisting of quark matter only. The linear $\sigma$-model is well
suited to consider the chiral symmetry breaking patterns of strong interactions
\cite{Lenaghan:2000ey}. The quark-quark interaction is
mediated by the exchange of meson fields. These interactions are conceptually
different from the NJL model which considers point coupling terms between
quarks. We consider scalar- and vector meson contributions to effectively
model the attractive and repulsive character of the strong interaction.  The
maximum masses of pure quark stars are calculated by solving the
Tolman-Oppenheimer-Volkoff (TOV) equations for different choices of the
parameters of the quark-meson model. We further investigate the parameter
space required for the existence of absolutely stable strange quark matter and
strange stars and confront these results with the new maximum mass limit for
compact stars. This model has been studied in \cite{Beisitzer:2014kea} for the
properties of quark matter under the conditions present in core-collapse
supernovae. However, the two solar mass constraint has not been investigated
so far. 


\section{The chiral quark-meson model}


The quark-meson model is based on the linear $\sigma$-model as discussed in
detail in \cite{Schaefer:2008hk,Lenaghan:2000ey} and couples mesons and quarks
by utilizing chiral symmetry. The mesonic contribution in this model is given by
\begin{eqnarray}
\label{lagrangiana} \nonumber
\mathcal{L}_{M}&=&tr(\partial_{\mu}\varphi)^{\dagger}(\partial^{\mu}\varphi)+
tr(\partial_{\mu} V)^{\dagger}(\partial^{\mu} V)\\ \nonumber
&-&\lambda_1[tr(\varphi^{\dagger}\varphi)]^2-\lambda_2
tr(\varphi^{\dagger}\varphi)^2\\ \nonumber
&-&m_0^2(tr(\varphi^{\dagger}\varphi))-m_v^2 (tr(V^{\dagger}V))\\ \label{mesonic}
&-&tr[\hat{H}(\varphi+\varphi^{\dagger})]+c\left(\det(\varphi^{\dagger})+\det(\varphi)\right)
\end{eqnarray}
for $SU(3)\times SU(3)$ chiral symmetry incorporating the scalar ($\varphi$) and vector
($V_\mu$) meson nonet.  Here, $m_v$ stands for the vacuum mass of the vector mesons
and $\lambda_1$, $\lambda_2$, $m_0$, and $c$ are the standard parameters of the linear
$\sigma$ model to be fixed below. The matrix $\hat{H}$ describes the explicit
breaking of chiral symmetry. 
The quarks couple to the meson fields via Yukawa-type interaction terms
\begin{equation}
\mathcal{L}_{Q}=\bar{\Psi}\left(i\cancel{\partial}
-g_\varphi\varphi-g_v\gamma^\mu V_\mu\right)\Psi 
\end{equation}
with the couplings strengths $g_\varphi$ and $g_v$ for scalar and vector
mesons, respectively.  Both contributions are forming the SU(3) Lagrangian $\mathcal{L}=\mathcal{L}_{M}+\mathcal{L}_{Q}$ of
the chiral quark-meson model.

In the mean-field approximation,  the matrix $\varphi$ consists of just the scalar
nonstrange field $\sigma_n$ and the strange field $\sigma_s$: 
\begin{equation}
\varphi=\frac{1}{\sqrt{2}}
\left(\begin{array}{ccc}
\frac{\sigma_n}{\sqrt{2}}&0&0\\
0&\frac{\sigma_n}{\sqrt{2}}&0\\
0&0&\sigma_s\\
\end{array}\right)
\end{equation}
and the vector fields are described by 
\begin{equation}
V=\frac{1}{\sqrt{2}}
\left(\begin{array}{ccc}
\frac{\omega+\rho}{\sqrt{2}}&0 &0\\
0 &\frac{\omega-\rho}{\sqrt{2}}&0\\
0&0&\phi\\
\end{array}\right)
\end{equation}
where $\rho$ stands for the zeroth component of the isovector vector field,
$\omega$ for the nonstrange vector field and $\phi$ for the strange
vector field, assuming ideal mixing.  Note that the spatial components of
$V_\mu$ vanish in the mean-field approximation in the static case to be
considered here\footnote{For an extension considering $\sigma_u$ and $\sigma_d$ seperately, see \cite{Stiele:2013pma}}. The Lagrangian for the quark fields then reads:
\begin{eqnarray} \nonumber
\mathcal{L}_{F_{n,s}}&=&\bar{\Psi}_n\left(i\cancel{\partial}
-g_{\omega}\gamma^0\omega-g_{\rho}\vec{\tau}\gamma^0\rho-g_n\sigma_n\right)\Psi_n\\ 
&+&\bar{\Psi}_s\left(i\cancel{\partial}-g_s \sigma_s-g_{\phi}\gamma^0\phi\right)\Psi_s
\end{eqnarray}
The indices $n$ and $s$ denote the nonstrange and strange quark
contributions. The quark fields couple to the scalar- and vector meson
fields $\sigma_n$, $\sigma_s$, $\omega$, $\rho$, and $\phi$ with the
respective coupling strength $g_i$, which are related by SU(3) flavor
symmetry to one overall coupling constant for the scalar meson $g_\varphi$ and
to another one for the vector coupling constant $g_v$.

Considering stationary fields in mean field approximation, the derivative
terms in $\mathcal{L}_M$ vanish and the Lagrangian is given by
\begin{eqnarray} \label{meinereiner} \nonumber
\mathcal{L}&=&\mathcal{L}_{M}+\mathcal{L}_Q\\ \nonumber
&=&\frac{1}{2}\left(m_{\omega}^2\omega^2+m_{\rho}^2\rho^2+m_{\phi}^2\phi^2\right)\\ \nonumber
&-&\frac{\lambda_1}{4}(\sigma_n^2+\sigma_s^2)^2-\frac{\lambda_2}{4}(\sigma_n^4+\sigma_s^4)\\
\nonumber 
&-&\frac{m_0^2}{2}(\sigma_n^2+\sigma_s^2)+\sqrt{2} \sigma_n^2\sigma_s
c+h_n\sigma_n+h_s\sigma_s-B\\ \nonumber 
&+&\bar{\Psi}_n\left(i\cancel{\partial}-g_{\omega}\gamma^0\omega
-g_{\rho}\vec{\tau}\gamma^0\rho-g_n\sigma_n\right)\Psi_n\\ 
&+&\bar{\Psi}_s\left(i\cancel{\partial}-g_s \sigma_s-g_{\phi}\gamma^0\phi\right)\Psi_s 
\end{eqnarray}
Here, a vacuum energy term $B$ has been introduced in addition, see the discussion
in \cite{Schertler:1997za,Schertler:2000xq,Pagliara:2007ph}. The electrons
will be treated as a free noninteracting Fermi-gas. The potential of the
Lagrangian (\ref{meinereiner}) then reads
\begin{eqnarray} 
\label{potV} \nonumber
\mathcal{V}&=&-\frac{1}{2}\left(m_{\omega}^2\omega^2+m_{\rho}^2\rho^2+m_{\phi}^2\phi^2\right)\\
\nonumber 
&+&\frac{\lambda_1}{4}(\sigma_n^2+\sigma_s^2)^2+\frac{\lambda_2}{4}(\sigma_n^4+\sigma_s^4)\\ 
&+&\frac{m_0^2}{2}(\sigma_n^2+\sigma_s^2)-\frac{2\sigma_n^2\sigma_s}{\sqrt{2}}\cdot
c-h_n\sigma_n-h_s\sigma_s+B 
\end{eqnarray}
for the meson fields. 


\section{The equation of state}
\label{eos}


The grand canonical potential is related to the partition function via
\begin{equation} \label{grandpot}
\Omega=-\frac{ln\mathcal{Z}}{\beta}=-p.
\end{equation}
The partition function $\mathcal{Z}$ can be computed by a Feynman path
integral over the quark fields. Performing the integration in mean field
approximation, the classical stationary mesonic background fields can be
replaced by their non-vanishing vacuum expectation values:
\begin{equation}\label{pfint}
\mathcal{Z}=\int \prod_a
\mathcal{D}\sigma_a\mathcal{D}\pi_a\int\mathcal{D}\bar{\Psi}\mathcal{D}\Psi 
e^{\left(\int_0^{\beta}d\tau \int_V d^3\vec{r}(\mathcal{L}+\bar{\Psi}\gamma^0\mu\Psi)\right)}
\end{equation}
and one arrives at
\begin{equation}\label{dumbledore}
\Omega=\mathcal{V}-\frac{3}{\pi^2\beta}\int^{\infty}_0 k^2dk \cdot \mathcal{M}
\end{equation}
where $\mathcal{V}$ is the potential given in eq.~(\ref{potV}) 
and the shorthand notation for $\mathcal{M}$ is
\begin{eqnarray}\nonumber
\mathcal{M}&=&\ln\left(1+e^{\frac{-E_u+\mu_u-g_{\omega}\omega-g_{\rho}\rho}{T}}\right)
+\ln\left(1+e^{\frac{-E_u-\mu_u+g_{\omega}\omega+g_{\rho}\rho}{T}}\right)\\ 
\nonumber
&+&\ln\left(1+e^{\frac{-E_d+\mu_d-g_{\omega}\omega+g_{\rho}\rho}{T}}\right)
+\ln\left(1+e^{\frac{-E_d-\mu_d+g_{\omega}\omega-g_{\rho}\rho}{T}}\right)\\ 
\label{ehmm}
&+&\ln\left(1+e^{\frac{-E_s+\mu_s-g_{\phi}\phi}{T}}\right)+
\ln\left(1+e^{\frac{-E_s-\mu_s+g_{\phi}\phi}{T}}\right)
\end{eqnarray}

where the flavor dependent energy is
\begin{equation}
E_{f}=\sqrt{k_{f}^2+(g_{f}\sigma_{f})^2}
\end{equation}
Compact star matter can be treated in the zero temperature limit $T
\rightarrow 0$. The equations of motion of the five meson fields are
given by minimizing the thermodynamic potential:
\begin{equation} \label{derivatives}
\frac{\partial \Omega}{\partial \sigma_n}=\frac{\partial
  \Omega}{\partial\sigma_s} 
=\frac{\partial \Omega}{\partial \omega}=\frac{\partial \Omega}{\partial \rho}
=\frac{\partial \Omega}{\partial \phi}\overset{!}{=}0 
\end{equation}
For the scalar mesons one finds the gap-equations
\begin{eqnarray}
\frac{\partial\Omega}{\partial\sigma_n}&=&\lambda_1\sigma_n(\sigma_n^2+\sigma_s^2)
+\frac{\lambda_2}{2}\sigma_n^3+m_{\sigma_n}^2\sigma_n-h_n\\
\nonumber 
&+&\frac{3g_n^2\sigma_n}{\pi^2}\left(\int_0^{k_{F}^u}\frac{dk\cdot
    k^2}{E_u}+\int_0^{k_{F}^d}\frac{dk\cdot k^2}{E_d}\right) = 0
\end{eqnarray}
and
\begin{eqnarray}
\frac{\partial\Omega}{\partial\sigma_s}&=&\lambda_1\sigma_s(\sigma_n^2+\sigma_s^2)
+\lambda_2\sigma_s^3+m_{\sigma_s}^2\sigma_s-h_s\\
\nonumber 
&+&\frac{3g_s^2\sigma_s}{\pi^2}\int_0^{k_{F}^s}\frac{dk\cdot k^2}{E_s} = 0
\end{eqnarray}
and for the vector fields 
\begin{eqnarray}
\frac{\partial\Omega}{\partial\omega}&=&-m_{\omega}^2\omega+ 
\frac{3g_{\omega}}{\pi^2}\left(\int_0^{k_{F}^u}dk\cdot
  k^2+\int_0^{k_{F}^d}dk\cdot k^2\right) = 0 \qquad \\  
\frac{\partial\Omega}{\partial\rho}&=&-m_{\rho}^2\rho+\frac{3g_{\rho}}{\pi^2}
\left(\int_0^{k_{F}^u}dk\cdot k^2-\int_0^{k_{F}^d}dk\cdot k^2\right) = 0\\ 
\frac{\partial\Omega}{\partial\phi}& = &
-m_{\phi}^2\phi+\frac{3g_{\phi}}{\pi^2}\int_0^{k_{F}^s}dk\cdot k^2 = 0 
\end{eqnarray}

and the respective Fermi momenta are
\begin{eqnarray}\label{ku}
 k_u&=&\sqrt{\left(\mu_q-\frac{2\mu_e}{3}-g_\omega\omega-g_{\rho}\rho\right)^2-(g_n \sigma_n)^2}\\ \label{kd}
 k_d&=&\sqrt{\left(\mu_q+\frac{\mu_e}{3}-g_\omega\omega+g_{\rho}\rho\right)^2-(g_n \sigma_n)^2}\\ \label{ks}
 k_s&=&\sqrt{\left(\mu_q+\frac{\mu_e}{3}-g_\phi\phi\right)^2-(g_s \sigma_s)^2}
\end{eqnarray}
 to guarantee charge neutrality through the electron chemical potential $\mu_e$.
Note that the source terms for the vector fields are given by the vector
number densities, the isovector number densities and the strange number
densities due to SU(3) flavor symmetry.

The coupled equations of motion of the meson fields have to be solved
self-consistently. Using eq.~(\ref{grandpot}) and the relation
\begin{equation}\label{ergden}
\epsilon=\Omega-\sum_i \mu_i n_i
\end{equation}
the resulting field values then determine the EoS, which serves as an input to
solve the TOV-equations:
\begin{eqnarray} 
\frac{dm}{dr}&=&\frac{4\pi r^2 \epsilon(r)}{c^2}
\label{tov_1} \\
\frac{dp}{dr}&=&-\frac{G \epsilon(r)m(r)}{(cr)^2}\\
&&\times\left(1+\frac{\rho(r)}{\epsilon(r)}\right)\left(1+\frac{4\pi
    r^3p(r)}{m(r)c^2}\right)
\left(1-\frac{2G m(r)}{c^2r}\right)^{-1} \quad
\label{tov_12} 
\end{eqnarray}
for the mass radius relation of compact stars. 


\subsection{Parameter range}
\label{pr-gag}

Within the mean field approximation the mesonic fields will be treated as
classical background fields. In the vacuum they will be replaced by their
vacuum expectation values (VEV).
In the scalar sector they are determined by the weak decay constants
\begin{eqnarray}
<\sigma_n>=f_{\pi}=92.4 \text{ MeV}\\ 
<\sigma_s>=\frac{2f_K-f_{\pi}}{\sqrt{2}}=94.47 \text{ MeV} 
\end{eqnarray}
with $f_K=159.8/\sqrt{2}$~MeV. The model incorporates four free parameters. The constituent quark mass will
be varied in a range around $m_q = 100$~MeV to 400~MeV. It determines the
scalar coupling for the nonstrange (\ref{eq:mqquark}) and strange (\ref{mqs})
condensates via the Goldberger-Treiman-relation
\begin{eqnarray}\label{eq:mqquark}
g_{u,d}=g_n=\frac{m_q}{f_{\pi}}
\end{eqnarray}
and from SU(3) symmetry
\begin{eqnarray}\label{mqs}
g_{s}=g_{n}\cdot\sqrt{2}\;.
\end{eqnarray}
The vector coupling is independent on the constituent quark mass, it will be
varied in a scale similar to the one of the scalar coupling, $g_{\omega}\sim
g_n$, to study its influences in an appropriate range \cite{Beisitzer:2014kea}. The coupling constant of
the $\phi$-meson is fixed again by SU(3) symmetry:
\begin{eqnarray}\label{eq:mqquarkss}
g_{\omega}=g_{\rho}=\frac{g_{\phi}}{\sqrt{2}}
\end{eqnarray}
To fix the remaining parameters of the scalar mesons, the following masses 
have been adopted:
\begin{eqnarray} \label{mpi}
m_{\pi}=138 \text{ MeV}\\ 
m_{K}=496 \text{ MeV}\\ 
m_{\eta}=547.5 \text{ MeV}\\ 
m_{\eta'}=957.78 \text{ MeV} 
\end{eqnarray}
The experimentally not well determined mass of the $\sigma$-meson could cover a
range from $m_{\sigma}=400$~MeV to 1000~MeV.  
The explicit symmetry breaking terms are defined via the
Gell-Mann-Oakes-Renner relation \cite{Lenaghan:2000ey} as
follows
\begin{eqnarray}\label{explicit}
h_n&=&f_{\pi}m_{\pi}^2\\ \label{explicit2}
h_s&=&\sqrt{2}f_Km_K^2-\frac{h_n}{\sqrt{2}} 
\end{eqnarray}
The parameter $\lambda_1$ has to be determined via the relation
\cite{Lenaghan:2000ey}
\begin{eqnarray} \label{hans}
m_{\sigma}^2(m^2,\lambda_1) \leftrightarrow
m_{\sigma}^2(m^2(\lambda_1),\lambda_1) 
\leftrightarrow m_{\sigma}^2(\lambda_1)
\end{eqnarray}
and $\lambda_2$ is given by
\begin{eqnarray}\label{l2}
\lambda_2&=&\frac{3m_K^2(2f_K-f_{\pi})-m_{\pi}^2(2f_K+f_{\pi})}
{(3f_{\pi}^2+8f_K(f_K-f_{\pi}))(f_K-f_{\pi})} \nonumber \\
&-&\frac{2(m_{\eta'}^2+m_{\eta}^2)(f_K-f_{\pi})}{(3f_{\pi}^2+8f_K(f_K-f_{\pi}))(f_K-f_{\pi})}
\end{eqnarray}
The large mass of the $\eta'$-meson (which, as a Goldstone boson should be
nearly massless, see \cite{Parganlija:2010fz,Koch:1997ei}) will be implemented
by the axial anomaly term with the coupling strength $c$ determined by
\begin{eqnarray}
c=\frac{m_K^2-m_{\pi}^2}{f_K-f_{\pi}}-\lambda_2\cdot(2f_K-f_{\pi}).
\end{eqnarray}
The last free parameter is the vacuum pressure at vanishing chemical potential
$B$, which we choose to cover a range from $B^{1/4}=0$~MeV to 140~MeV. While
varying one parameter in the following, the other parameters will be held
fixed at some canonical values chosen to be $m_{\sigma}= 600$~MeV,
$m_q=300$~MeV, $g_{\omega}=2$ and $B^{1/4}=120$~MeV.


\section{Results}


\subsection{Variation of the vector coupling constant $g_{\omega}$}
\label{varofomega}

The vector coupling models the repulsive character of the strong interaction. 
Analogue to the scalar coupling, which is dependent
on the constituent quark mass, the vector coupling will be changed in a range
from $g_{\omega }= 1$ to 7.

\begin{figure}
\includegraphics[width=\columnwidth]{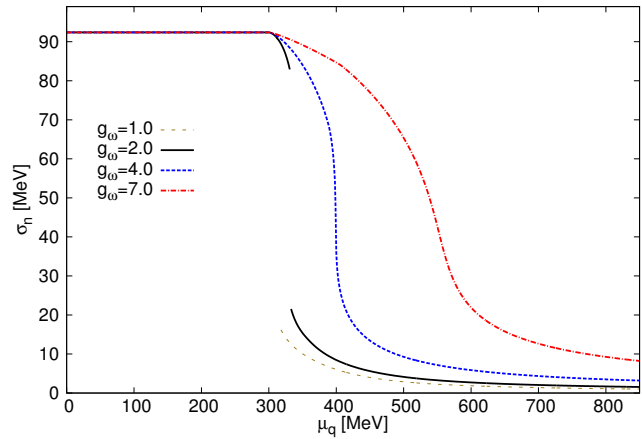}
\caption{The nonstrange scalar condensate for various values of the
    vector coupling constant $g_{\omega}$ at fixed values of $m_q=300$~MeV,
    $m_{\sigma}=600$~MeV and $B^{1/4}=120$~MeV.}
\label{fig:gw_fields}
\end{figure}

Figure \ref{fig:gw_fields} shows the nonstrange scalar condensates for various
values of $g_{\omega}$. There is a crossover for values of $g_{\omega} \gtrsim
4$. For values of $g_{\omega} \lesssim 4$ there is a jump in the scalar
condensate, which is present at smaller chemical potential $\mu_q$ for smaller
values of $g_{\omega}$. The chiral transition for $g_{\omega}=2.0$ takes place
at $\mu \sim 325$~MeV.  Because of explicit symmetry breaking, the condensate
does not vanish entirely in the chirally restored phase, see e.g.\ the
discussion in \cite{Lenaghan:2000ey}.

\begin{figure}
\includegraphics[width=\columnwidth]{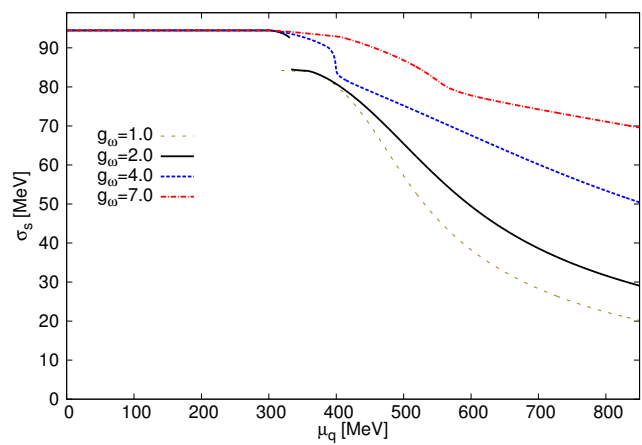}
\caption{The strange scalar condensates for various values of the
    vector coupling constant $g_{\omega}$ at fixed $m_q=300$~MeV,
    $m_{\sigma}=600$~MeV and $B^{1/4}=120$~MeV.}
\label{fig:ububu}
\end{figure}

Note that the jump is also visible for the strange condensate, which can be
seen in Fig.~\ref{fig:ububu}. Because of the larger mass of the strange
quark the jump is not that pronounced and the field changes rather smoothly
staying at larger values even in the chirally restored phase up to the largest considered 
values of the chemical potential.

An increase of the repulsive coupling $g_{\omega}$ provokes the scalar fields to increase too.
This is due to the substraction of the terms including the vector fields from the chemical 
potential $\mu_q$ in eqs.~(\ref{ku}), (\ref{kd}) and (\ref{ks}).
Larger vector field terms require larger chemical potential to compensate for their impact.

\begin{figure}
\includegraphics[width=\columnwidth]{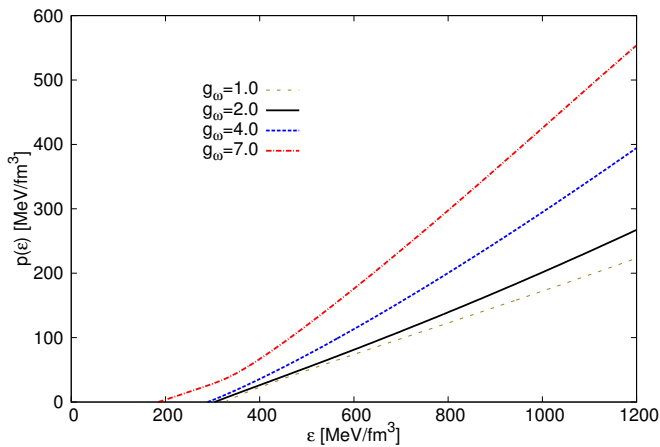}
\caption{The EoS for various values of the vector coupling constant
    $g_{\omega}$ at fixed $m_q=300$~MeV, $m_{\sigma}=600$~MeV and
    $B^{1/4}=120$~MeV.} 
\label{fig:gweos}
\end{figure}

The EoS for various values $g_{\omega}$ is plotted in
Fig.~\ref{fig:gweos}. For a larger value of the vector coupling constant, the
repulsion between quarks increases, the slope of the EoS increases and the EoS
becomes stiffer, i.e.\ quark matter requires more pressure to be compressed to
a given energy density.

\begin{figure} 
\includegraphics[width=\columnwidth]{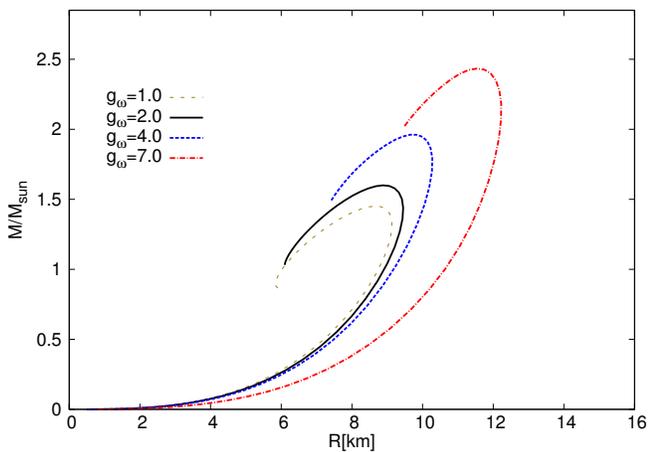}
\caption{The mass-radius relation for various values of the vector
    coupling constant $g_{\omega}$ at fixed $m_q=300$~MeV, $m_{\sigma}=600$~MeV
    and $B^{1/4}=120$~MeV.}
\label{fig:gwmass}
\end{figure}

The corresponding solutions of the TOV equations can be seen in
Fig.~\ref{fig:gwmass}. With a stiffer EoS the maximum mass increases. This
increase in the maximum mass is due to the repulsive character of the vector
mesons which gives a higher pressure at a given energy density (see
Fig.~\ref{fig:gweos}) and therefore quark matter is able to stabilize more
mass against the pull of gravity. With the choice of $g_{\omega}=4.0$ and the
other parameters held fixed at $m_q=300$~MeV, $m_{\sigma}=600$~MeV and
$B^{1/4}=120$~MeV one reaches a maximum mass of $\sim 2.1M_{\odot}$ at a
radius of 10~km. Smaller values of the vector coupling constant result in
compact star configurations with a maximum mass smaller than $2M_\odot$ which is in
conflict with observations. Hence, the vector-like interactions between quarks
are necessary in the model used to achieve the $2M_{\odot}$ mass limit in a
physically reasonable range of our parameters.


\subsection{Variation of the constituent quark mass}
\label{mukuh}

Raising the value of the constituent quark-mass increases the scalar coupling,
see equations (\ref{eq:mqquark}) and (\ref{mqs}).  The values of the scalar
condensate have been checked with the results of \cite{Schaefer:2008hk} and
found to be in accordance.

We find that a crossover transition is present for values of $\mu_q \lesssim
300$~MeV, whereas for values $\mu_q \geq 300$~MeV a first order phase
transition emerges, and the chiral condensate jumps. For even higher values,
the strength of the first order chiral phase transition increases.

\begin{figure} 
\includegraphics[width=\columnwidth]{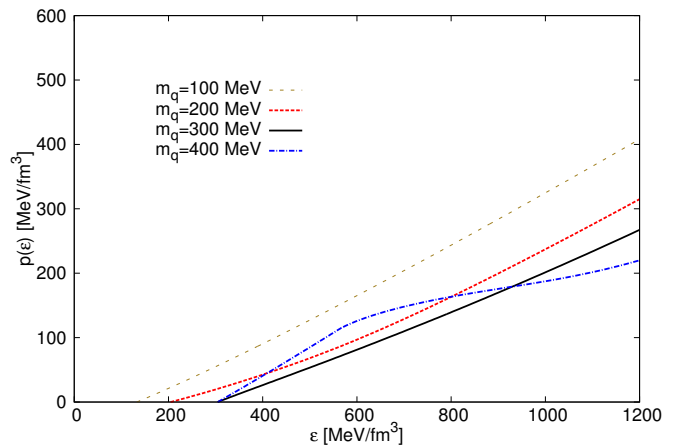}
\caption{The EoS while varying $m_q$ with fixed $g_{\omega}=2.0$,
    $m_{\sigma}=600$~MeV and $B^{1/4}=120$~MeV.}
\label{fig:mqeos}
\end{figure}

\begin{figure} 
\includegraphics[width=\columnwidth]{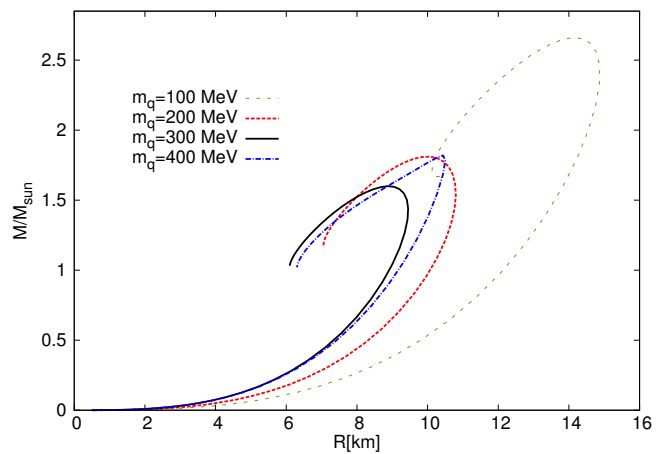}
\caption{The mass-radius relation for various values of $m_q$ at fixed
    $g_{\omega}=2.0$, $m_{\sigma}=600$~MeV and $B^{1/4}=120$~MeV}
\label{fig:massessschatz}
\end{figure}

The EoS shown in Fig.~\ref{fig:mqeos} displays a softening for increasing
values of $m_q$, leading to smaller maximum masses of the compact star
configuration, as can be seen in Fig.~\ref{fig:massessschatz}.
The trend of the curve of the EoS for $m_q \geq 300$~MeV shows a slightly different behaviour.
For $m_q=100$~MeV the $2M_{\odot}$ -limit is exceeded. In this case 
the scalar condensates would exhibit a smooth crossover-like behavior as a
function of the chemical potential.


\subsection{Variation of the $\sigma$-Meson mass}

We find that with increasing $\sigma$ meson mass $m_{\sigma}$ the phase
transition becomes a crossover ($m_{\sigma}\geq 800$~MeV), whereas values of 
$m_{\sigma} \leq 800$~MeV lead to first order
phase transitions at $\mu_q \simeq 300$~MeV. For $m_{\sigma}=600$~MeV the
first order phase transition takes place at $\mu_q \sim 330$~MeV. The behavior
of the nonstrange and strange condensates is found to be similar to the discussion in section
\ref{varofomega}. 

\begin{figure} 
\includegraphics[width=\columnwidth]{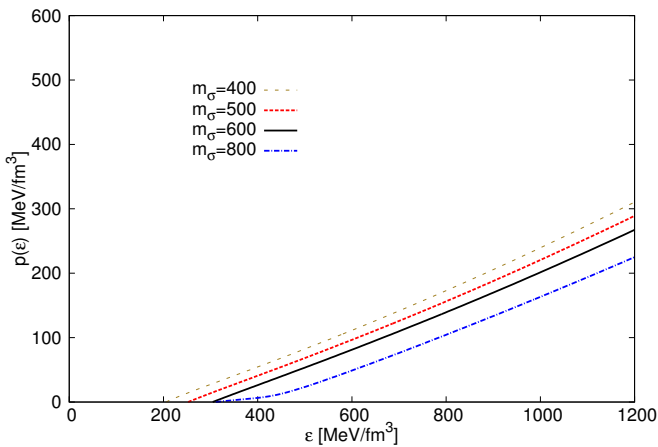}
\caption{The EoS for various values of $m_{\sigma}$ at fixed
    $g_{\omega}=2.0$, $m_{q}=300$~MeV and $B^{1/4}=120$~MeV.}
\label{fig:mseos}
\end{figure}

\begin{figure} 
\includegraphics[width=\columnwidth]{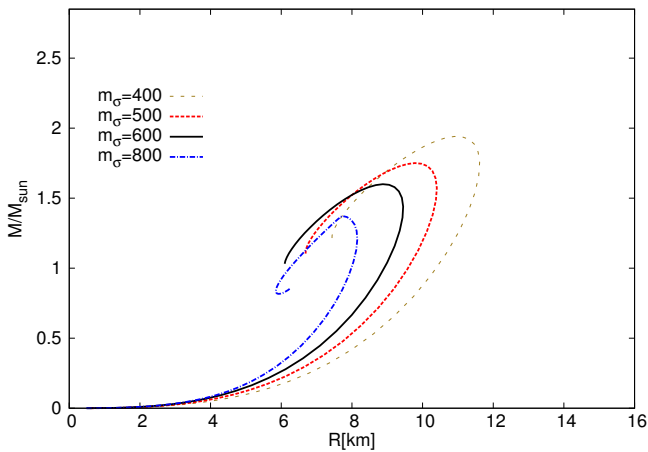}
\caption{The mass-radius relation for various values of $m_{\sigma}$
    at fixed $g_{\omega}=2.0$, $m_{q}=300$~MeV and $B^{1/4}=120$~MeV.}
\label{fig:msmass}
\end{figure}

The resulting EoS is shown in Fig.~\ref{fig:mseos}. The EoS softens with
increasing scalar meson mass $m_{\sigma}$. Hence, one expects a mass-radius
relation which is located at smaller values of the mass and radius for
increasing the $\sigma$ meson mass, which can clearly be observed in
Fig.~\ref{fig:msmass}. So a smaller value of $m_{\sigma}$ leads to higher
maximum masses of the compact star.

The mass of the sigma meson $m_{\sigma}$ is directly related to the parameters
$\lambda_1$ and $\lambda_2$, according to (\ref{hans}) and (\ref{l2}). These
parameters are, among others in the SU(3) case, mainly responsible for the
potential depth for spontaneous symmetry breaking. The impact of the sigma
meson mass on the chiral condensate and the EoS is nontrivial. A simple
explanation might be that for a higher mass of the scalar meson more energy is
needed to overcome the barrier and reach the second minimum of the potential,
which leads also to a shift of the chiral phase transition to a larger
chemical potential $\mu_q$ and a crossover-like behavior of the chiral
condensates at high densities.


\subsection{Variation of the vacuum constant}

\begin{figure} 
\includegraphics[width=\columnwidth]{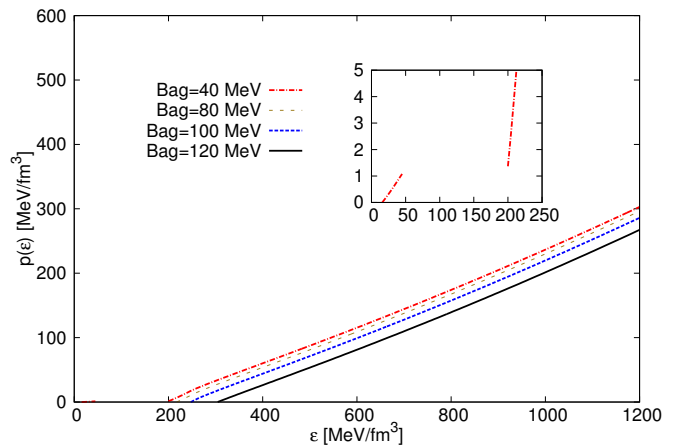}
\caption{The equation of state for various values of the vacuum constant
  $B^{1/4}$ at fixed values of $m_\sigma=600$~MeV, $g_\omega=2.0$
  and $m_q=300$~MeV.}
\label{fig:bageos}
\end{figure}

Figure~\ref{fig:bageos} shows the equation of state for various values of the
vacuum constant $B^{1/4}$. The equation of state does not change significantly
when varying the vacuum constant. The vacuum constant $B$ drops out 
in the equations of motion (\ref{derivatives}) so that it does not affect the 
values of the meson fields. 

\begin{figure} 
\includegraphics[width=\columnwidth]{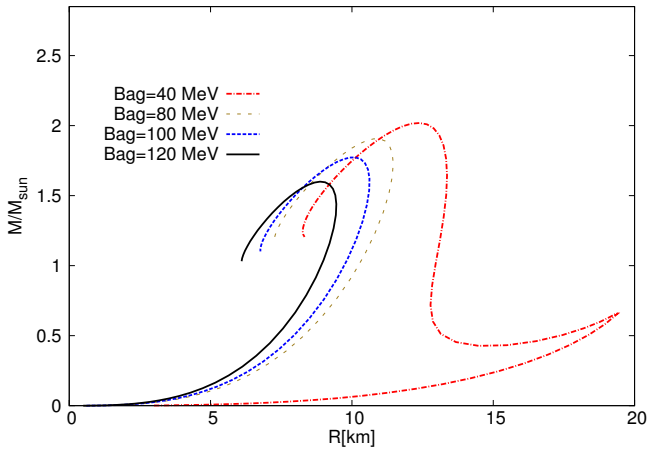}
\caption{The mass-radius relation for various values of the vacuum constant
  $B^{1/4}$ with fixed values of $m_{\sigma}=600$MeV, $g_{\omega}=2.0$ and
  $m_{q}=300$~MeV. The maximum mass increases significantly for a lower value
  of the vacuum constant $B$.}
\label{fig:bagmass}
\end{figure}

The slopes of the curves are very similar to each other. The pressure
vanishes at lower energy densities for a smaller vacuum constant. 
Note that the curve for the vacuum constant $B^{1/4}=40$~MeV is not simply shifted 
but the jump in the energy density is in this case at a small, but non-vanishing value of the pressure. This can be observed in the inlay of Fig.~\ref{fig:bageos}. 
This property leads to an additional stable branch in the mass-radius relation shown in Fig.~\ref{fig:bagmass}.
For $B^{1/4}=40$~MeV there are two stable branches. One up to 0.6~$M_{\odot}$ at a radius 
of $\simeq 19$~km and a second branch between $\simeq 14$~km and $\simeq 12$~km radius and up to a maximum mass of $\sim 2 M_{\odot}$.
These so called twin star solutions \cite{Schertler:1997za,Schertler:2000xq} are beyond the scope
of this article and are discussed in more detail in \cite{Schertler:1997za,Glendenning:1998ag,Hanauske:2001tc,Alvarez-Castillo:2013cxa,Alford:2014dva,Alford:2015dpa,Banik:2002kc,Blaschke:2015uva} 
and in the forthcoming publication \cite{Zacchi:2015wak}.
See \cite{Glendenning:1998ag} for a detailled discussion on stability.

The mass-radius relation depicted in Fig.~\ref{fig:bagmass} demonstrates
that larger maximum masses can be achieved for smaller values of the vacuum
constant $B$. We point out that the standard choice of the vacuum pressure in
refs.~\cite{Schertler:1997za,Schertler:2000xq} was a value of
$B^{1/4}=120$~MeV.  For this value the resulting mass-radius relation has a
maximum mass around $1.6M_{\odot}$ at a radius of around 9~km, when fixing the
other parameters of the model at their standard values. The vacuum constant has a
strong influence on the mass-radius relation, and the maximum mass of a
compact star can easily reach values of $2 M_{\odot}$. Choosing
$B^{1/4}=40$~MeV the resulting compact star configurations have a maximum mass
of $\sim 2 M_{\odot}$ at a radius of 12~km.


\subsection{Stability considerations}

The properties of quark matter have to fulfill certain conditions in order to 
allow for the possibility of pure quark matter stars studied
above. 

Normal matter, consisting of ordinary nuclei, is stable on cosmological timescales, 
so it does not decay to quark matter with its quark constituents
of up-quarks and down-quarks. This observation requires that 
two flavor quark matter can not be more stable than ordinary nuclear matter,
meaning that the energy per baryon has to be higher than the one of the most stable
known element in nature: $^{56}$Fe. We adopt a value of energy per baryon of 
$E/A=930$~MeV for nuclei and add a $4
$~MeV correction due to surface effects of lumps of quark matter balls as
discussed in \cite{Farhi:1984qu}. Hence the critical condition for two-flavor
quark matter reads
\begin{equation}\label{lotharmatthaeus}
\frac{E}{3A}\bigg|_{p=0}=\frac{\epsilon}{n_q}\bigg|_{p=0}\gtrsim 311 \text{ MeV}
\end{equation}
This condition guarantees the stability of atomic nuclei, meaning that atomic
nuclei do not dissolve into their constituent quarks.

On the other hand three-flavor quark matter, i.e.\ quark matter consisting of
up-, down- and strange quarks, could be more stable than ordinary nuclei which
is the Bodmer-Witten hypothesis \cite{Bodmer:1971we, Witten:1984rs}.  Ordinary
nuclear matter can not decay to strange quark matter, as there is a barrier
between these two states of matter due to strangeness conversion via weak
interactions with a corresponding conversion timescale much longer than the
age of the universe. The presence of the new degree of freedom, the strange
quark, in quark matter lowers the overall energy per baryon, so that this
state could be energetically favorable compared to nuclear matter. Hence, the
condition for absolutely stable strange quark matter of 
$E/A<930$~MeV can be recast in the form 
\begin{equation}\label{sfe3ede}
\frac{E}{3A}\bigg|_{p=0}=\frac{\epsilon}{n_q}\bigg|_{p=0}\leq 310 \text{ MeV}
\end{equation}
In the following we will denote the physical condition for two-flavor quark
matter, eq.~(\ref{lotharmatthaeus}), the two-flavor condition (or two-flavor
line in the plots) and the one for three-flavor quark matter,
eq.~(\ref{sfe3ede}), the three-flavor condition (or three-flavor line in the
plots).

\begin{figure} 
\includegraphics[width=\columnwidth]{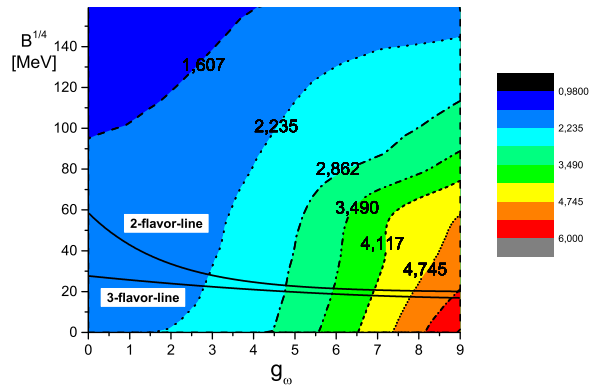}
\caption{Maximum masses for three-flavor pure quark stars in the plane of the
  vacuum constant $B$ and the vector coupling constant $g_\omega$. The
  two-flavor and three-flavor lines delineate the region for hybrid stars (above the
  two-flavor line) and pure quark stars (below the three-flavor line). Pure quark
  star configurations do not appear for the parameters chosen
  ($m_\sigma=600$~MeV and $m_q=300$~MeV).}
\label{fig:sc1}
\end{figure}

\begin{figure}  
\includegraphics[width=\columnwidth]{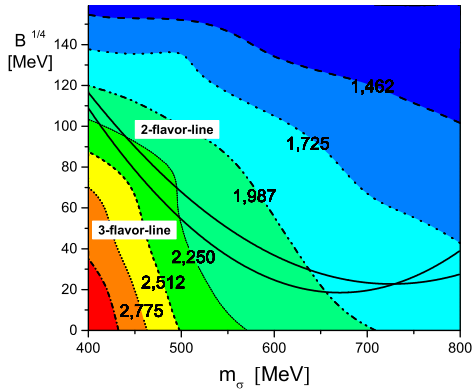}
\caption{Maximum masses for three-flavor pure quark stars in the plane of the
  vacuum constant $B^{1/4}$ and the sigma meson mass $m_\sigma$. The two-flavor
  and three-flavor lines delineate the region for hybrid stars (above the two-flavor
  line) and pure quark stars (below the three-flavor line). Pure quark star
  configurations are possible for a small window at large values of the sigma
  meson mass $m_\sigma$ (here $g_\omega =2$ and $m_q=300$~MeV are held
  fixed).}
\label{fig:sc2}
\end{figure}

Figures~\ref{fig:sc1} and \ref{fig:sc2} depict the maximum masses in
dependence of the parameters $B^{1/4}$, $m_{\sigma}$ and $g_{\omega}$. In
general, smaller values of the vacuum constant $B$ and higher values for
the vector coupling constant $g_{\omega}$ lead to higher maximum masses.

Figure \ref{fig:sc1} shows as a contour plot the dependencies of the
maximum mass of pure quark star configurations on the vacuum
constant $B^{1/4}$ and the vector coupling constant $g_{\omega}$. For $B^{1/4}=120$~MeV and
$g_{\omega}=2.0$ one finds a maximum mass of about $1.6 M_{\odot}$, which can
be cross-checked with Figs.~\ref{fig:gwmass}, ~\ref{fig:massessschatz}, ~\ref{fig:msmass} and ~\ref{fig:bagmass}.

The two-flavor line indicates the two-flavor constraint
(\ref{lotharmatthaeus}) and the three-flavor line the three-flavor constraint
(\ref{sfe3ede}). In the area above the two-flavor line the condition
(\ref{lotharmatthaeus}) is fulfilled, i.e.\ normal matter can not decay to
two-flavor quark matter as observed in nature. For a vanishing vector
repulsion the vacuum constant has to be larger than $B^{1/4}>60$~MeV in order
for quark matter to obey the two-flavor constraint. The critical value for
$B^{1/4}$ decreases towards $B^{1/4}=20$~MeV with increasing vector coupling
constant. Hybrid star configurations are located in the
parameter range above the two-flavor line.

In the area below the three-flavor line the condition (\ref{sfe3ede}) is
fulfilled. Within the chosen parameters the three-flavor line is nearly
independent on the vector coupling constant $g_\omega$ being between
$B^{1/4}\sim 28$~MeV for a vanishing vector coupling constant and $B^{1/4}\sim
20$~MeV for $g_{\omega}=9.0$. The vacuum pressure dictates at which energy
density $\epsilon_0$ the pressure vanishes, which determines via the
Hugenholtz-van Hove theorem the binding energy of quark matter:
$E_B/A=\mu_B=\epsilon_0/n_B$. As, there is no region, where both conditions
(\ref{lotharmatthaeus}) and (\ref{sfe3ede}) are fulfilled simultaneously, no
pure quark star configurations emerge in the contour plot for the chosen
parameters of $m_\sigma=600$~MeV and $m_q=300$~MeV.

Figure \ref{fig:sc2} shows the maximum masses and the quark matter constraints
(\ref{lotharmatthaeus}) and (\ref{sfe3ede}) in the parameter plane of the
vacuum constant $B^{1/4}$ and the $\sigma$-meson mass $m_{\sigma}$.  The
two-flavor line starts at $B^{1/4}=115$~MeV for $m_{\sigma}=400$~MeV, with the
three-flavor line being slightly below the two-flavor line. Both lines
decrease in a similar manner with increasing mass of the $\sigma$-meson. At
$B^{1/4}=26$~MeV and $m_{\sigma}=725$~MeV both lines intersect and from that
point on the three-flavor line is above the two-flavor line. This means that
from there on, in a small parameter space (the small area enclosed by both
lines), pure quark star configurations are stable.  The corresponding highest
maximum mass for pure quark stars in Fig.~\ref{fig:sc2} with $g_{\omega}=2$ and $m_q=300$~MeV held fixed is located at
$m_{\sigma} \sim 725$~MeV and $B^{1/4} \sim 26$~MeV with a value of $\sim
1.8M_{\odot}$ which is not compatible with the recent pulsar mass
measurements.

A high mass of the sigma meson seems to be necessary in order to fulfill
the constraints for pure quark star configurations, i.e.\ equation
(\ref{lotharmatthaeus}) and (\ref{sfe3ede}), in contrast to the variation of the
vector coupling constant. A projection on the $g_{\omega}$-$m_{\sigma}$-plane
on the other hand (with a fixed value of $B^{1/4}=120$~MeV) leads to a
null result for pure quark star configurations due to the high value of the
vacuum constant. Only a small value of the vacuum constant leads to pure quark
star configurations.

\begin{figure}  
\includegraphics[width=\columnwidth]{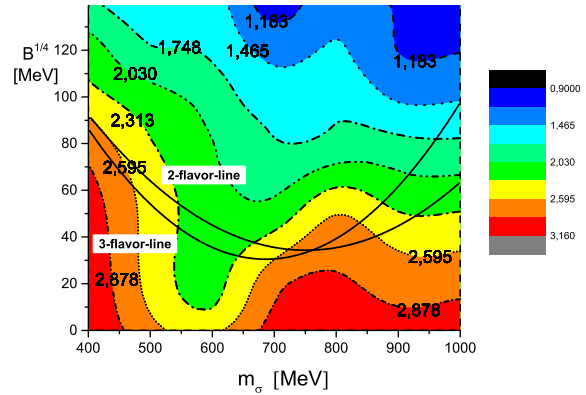}
\caption{Maximum masses and stability configurations for pure quark star
  configurations as in Fig.~\ref{fig:sc2} for a slightly larger vector
  coupling constant of $g_{\omega}=3.0$.}
\label{fig:sc3}
\end{figure}

As the two solar mass limit for pure quark stars with $g_{\omega}=2.0$ is
not reached, we increase the repulsive vector interaction $g_{\omega}$ 
in the following to a value of $g_{\omega}=3.0$. The resulting maximum masses
and lines of constraints are shown in Fig.~\ref{fig:sc3}. There is an
overall increase of the maximum mass as expected for a greater repulsive
interaction. Surprisingly, for $m_{\sigma} \geq 600$~MeV the maximum mass
increases with the $\sigma$ meson mass again, which can not be seen in Fig.~\ref{fig:sc2}.
This behavior indicates a switch in the dominance of the scalar- and vector
field contributions to the stiffness of the equation of state. Above a certain
value, around $g_{\omega} \sim 2.75$, the repulsive fields gain on their
influence on the maximum masses compared to the attractive scalar fields. A
combination of $g_{\omega}\gtrsim 2.75$ and $m_{\sigma}\geq 600$~MeV leads
then to higher maximum masses with increasing $\sigma$ meson mass instead. 

The lines of the two constraints intersect at $B^{1/4} \sim 32$~MeV and
$m_{\sigma} \sim 760$~MeV. From this point on pure quark star configurations
are realized between the two lines of constraint as discussed above. Note that in Fig.~\ref{fig:sc3}
results are shown for $\sigma$ meson masses of up to $m_\sigma
=1000$~MeV.  At the intersecting point at $B^{1/4} \sim 32$~MeV and
$m_{\sigma}\sim 760$~MeV the maximum mass of the quark star would be $\sim
2.7M_{\odot}$, being smaller for larger values of the $\sigma$ meson
mass $m_{\sigma}$ and larger values of the vacuum constant. A pure quark star
with a maximum mass of $\sim 2.0M_{\odot}$ can be found at 
a vacuum constant of $B^{1/4}=70$~MeV for $\sigma$ meson masses
between $900\mbox{ MeV} \leq m_{\sigma} \leq 1000$~MeV.


\subsection{Comparison with other models}
\label{quarkstable}

In the following we compare our findings with other approaches for studying
pure quark stars, i.e.\ selfbound strange stars. For hybrid stars not
discussed here a low-density hadronic equation of state needs to be augmented
for a thorough discussion which is beyond the scope of the present
investigation, but will be adressed in \cite{Zacchi:2015wak}. 

Coelho et al.\ \cite{Coelho:2010xw} use a SU(2)-flavor symmetric NJL model with
a repulsive vector coupling. For a large vector coupling their EoS stiffens
like in our model calculations leading to higher masses at given radii.

Weissenborn et al.\ \cite{Weissenborn:2011qu} use an extended quark bag
model. Strange stars can reach maximum masses beyond $2M_{\odot}$ in their
work if additional terms compared to the standard MIT bag model are
introduced, either in the form of some effective interaction motivated from
one-gluon exchange or from a large gap motivated from
color-superconductivity. They found a maximum mass for a pure quark star to be
at $2.54M_{\odot}$, which is in the same order of magnitude as in this work.

In the work by Torres and Menezes \cite{Torres:2012xv} pure quark stars would
have maximum masses of $2.29M_{\odot}$. They use a model, where the quark
masses are assumed to have a certain given density dependence causing a
stiffening of the EoS compared to the standard MIT bag model.


\section{Summary}


In this work a chiral Quark-Meson model in SU(3)-flavor symmetry has been
studied for the description of compact stars consisting of pure quark matter.
The thermodynamical properties have been calculated via the grand potential in
the zero temperature limit. The gap equations were solved selfconsistently to
determine the EoS, which serves as an input to solve the TOV equations and
compute the mass-radius relations. The EoS and the dependence on the four free
parameters of the model, the vector coupling constant, the constituent quark
mass, the sigma meson, and the vacuum constant, were systematically
investigated.

The variation of the vector coupling constant showed the highest impact on the
EoS. The higher its value, the stiffer the EoS, leading to maximum masses in
excess of $2M_{\odot}$. The dependence of the EoS with the constituent quark
mass $m_q$ in vacuum, which fixes the scalar coupling constant, is such that 
the smaller $m_q$, the stiffer the EoS.  The mass of the
$\sigma$-meson studied covers a range from 400 to 1000~MeV. For a smaller mass
of the $\sigma$-meson the EoS becomes stiffer. Finally, the vacuum constant
does not affect the values of the meson fields, it just shifts the energy
density at a given pressure. The EoS substantially stiffens when decreasing
the vacuum constant so that for small values of $B^{1/4}\lesssim 40$~MeV
maximum masses of up to $\geq 2M_{\odot}$ could easily be reached.

The stability of two-flavor and three-flavor quark matter have been checked
as well, to see whether or not the model parameter space allows for physically
reasonable quark matter properties in the SU(3)-flavor approach. The
$2 M_{\odot}$ limit set by the recently discovered millisecond pulsars PSR
J1614-2230 \cite{Demorest:2010bx} and PSR J0348+0432 \cite{Antoniadis:2013pzd}
could be reached. Having considered the stability constraints in the
equations (\ref{lotharmatthaeus}) and (\ref{sfe3ede}), most choices of the
parameter space were found to be hybrid stars. Nonetheless, pure quark star
configurations with $\sim 2M_{\odot}$ can be realized in a small physically
reasonable parameter range. A sizable additional repulsion mediated by the
exchange of vector mesons as well as a nonvanishing vacuum pressure seems to
be crucial to allow for maximum mass configurations of quark stars compatible
with the recent pulsar mass measurements.

\begin{acknowledgments}
  We thank Margit Maly for discussions during the initial stage of this
  project.  AZ was supported by the Friedrich-Ebert-Stiftung (FES).  RS has been
  supported by the Heidelberg Graduate School of Fundamental Physics (HGSFP)
  and through the Helmholtz Graduate School for Heavy-Ion Research (HGS-HIRe)
  and the Graduate Program for Hadron and Ion Research (GP-HIR).
\end{acknowledgments}

\bibliography{myliterature,all_new}
\bibliographystyle{apsrev4-1}

\end{document}